\documentstyle[twoside,fleqn,epsf,espcrc2]{article}
\newcommand{\link}{U}

\title{
       Two-loop Perturbative Quark Mass Renormalization from
       Large $\beta$ Monte Carlo
			}

\author{
        K.J.~Juge\address{Fermi National Accelerator Laboratory, 
                          P.O.~Box 500, Batavia, Illinois 60510}
			 }

\begin{document}

\begin{abstract}
We present the calculation of heavy Wilson quark mass renormalization constants from large beta Monte Carlo simulations. Simulations were performed at various beta larger than 9, each on several spatial lattice sizes to allow for an infinite volume extrapolation. We use twisted boundary conditions to suppress tunneling and work in Coulomb gauge with appropriate adjustments for the temporal links. The one-loop coefficient obtained from this method is in agreement with the analytical result and a preliminary result for the second order coefficient is reported.
\end{abstract}

\maketitle

\section{INTRODUCTION}
As actions for lattice calculations become more complicated, short distance calculations for normalizing the operators in the action become more and more difficult.  It is useful to have as many tools as possible to attack the challenge.  Several types of automated calculation of Feynman diagrams and nonperturbative methods have been developed.  Another alternative is the use of direct calculation of short distance quark--gluon Green functions by Monte Carlo methods\cite{dimm}.
Last year this method was applied to a determination of the additive mass renormalization for static quarks to third order\cite{lat99}. In this paper, we present a preliminary estimate of the mass renormalization for heavy Wilson quarks.

\section{SIMULATION}

There are several potential problems in naively performing simulations at large $\beta$ to extract perturbative information. One is the problem associated with infra-red divergences coming from the zero modes. The second is the phenomenon of tunneling between the various vacua of $SU(3)$. The third is finite volume effects, since the physical sizes of space-time used in the simulations are quite small. All of these effects may be removed or reduced by using twisted boundary conditions\cite{weisz} for the fields. These special boundary conditions remove the zero modes completely and suppresses tunneling for sufficiently large $\beta$\cite{lat99}. The remaining finite size effects can be removed by extrapolating to the infinite volume limit.

\subsection{Twisted Boundary Conditions}
We use a 3-way periodic twist (spatial directions only) for the gauge fields defined by,
\begin{eqnarray}
\link_\mu(x+L_j\hat{e}_j)&=&\Omega_j\link_\mu(x)\Omega^{-1}_j\\
\Omega_x\Omega_y&=&\eta\Omega_y\Omega_x 
\end{eqnarray}
where $\eta=e^{2\pi i/3}$ and $\Omega_z=\Omega_y\Omega^\dagger_x$. The $\Omega$'s are constant $SU(3)$ matrices.

\subsection{Gauge Fixing}
We must fix the gauge in order to extract energies from quark propagators. Our gauge fixing procedure is as follows:\\
(1) fix to Coulomb gauge, then\\
(2) fix the temporal links by maximizing,\\
$$tr\left[Z(t)\sum_{\vec{x}}U_t(\vec{x},t)Z(t+1)\right]$$
where $Z(t)$ are elements of the center of $SU(3)$.

\subsection{Parameters}
The following values of the coupling were chosen:
$\beta=9,10,12,15,20,40$ and $60$. Lattice sizes of $L_s=\{4,6,8,10,12,14,16\}$ and $L_t=16$ were used at each coupling to allow for an extrapolation to the infinite volume limit. In this preliminary study, we have accumulated $500\sim2000$ configurations separated by 100 sweeps. The errors after binning did not show significant residual correlations.

%%%%%%%%%%%%%%%%%%%%%%%%%%%%%%%%%%
\section{STATIC QUARK SELF-ENERGY}
%%%%%%%%%%%%%%%%%%%%%%%%%%%%%%%%%%
We test our gauge-fixed configurations by calculating the first two coefficients of the static quark self-energy and compare to analytic\cite{heller} and previous high $\beta$ Monte-Carlo calculations\cite{lat99} where no gauge fixing was performed. We also compare results from the Polyakov loop and the static quark propagator. 

\subsection{Polyakov Loop}
%-------------------------
The tadpole-improved static quark self-energy (in a finite volume) was calculated from the measured value of the Polyakov loop according to,
\begin{equation}
E_0(L) = -\frac{1}{L} {\mathrm ln}\left[
\frac{1}{3}{\it Re Tr}\left\{\prod^{L}_{t=1} \frac{\link_t(t)}{u_0}\right\}\right]
\end{equation}
where $u_0$ is the mean link from the average plaquette. Our fitting procedure to extract the renormalization constants is as follows:
\begin{itemize}
\item[1.] Fit the measured values at fixed $L$ to a polynomial in $\alpha_V(q^\star)$ to obtain $c_1(L)$ and $c_2(L)$, {\it i.e.} $E_0(L)/\alpha = c_1(L) + \alpha c_2(L) + \alpha^2c_3(L)$.
\item[2.] Extrapolate the coefficient $c_1(L)$ using $c_1(L) = c_1 + X_1/L$ to get the one-loop coefficient, $c_1$.
\item[3.] Extrapolate $c_2(L)$ with $c_2(L) = c_2 + X_2/L + \beta_0X_1ln(L)/L$ where $\beta_0 = 11/2\pi$ to get the two loop coefficient, $c_2$.
\end{itemize}

The functional form and the coefficient of the logarithm comes from the anticipation of the (running) Coulomb interaction with the image charges with some effective distance $L/X_1$.

We also perform the fit in the reverse order, {\it i.e.} extrapolate to infinite volume at each $\beta$ and then fit to a cubic polynomial in $\alpha_V$. In the infinite volume extrapolation, we use $E_0(L) = E_0 + Y_1/L + \beta_0X_1ln(L)/L$ where $X_1$ is the coefficient from the previous fit. The results were in perfect agreement. The renormalized coupling in the V-scheme\cite{lepage}, $\alpha_V$, was evaluated at $q^\star=0.84a^{-1}$. 

\subsection{Static Quark Propagator}
%-----------------------------------
On the same set of lattices, we have measured the static quark propagator averaged over the spatial sites. These were then fit to single exponentials starting from timeslice 2 for which we get a good fit. The perturbative coefficients were extracted in the exact same manner as in the previous section. 

\subsection{Comparison}
%----------------------
We compare our results with the analytical calculations of Heller and 
Karsch\cite{heller} and with previous measurements\cite{lat99} of the first two coefficients in Table~\ref{comp}. Our measurements on gauge fixed configurations with both gauge invariant and covariant operators are in good agreement with those in the literature. These results give us confidence in our gauge fixing procedure.

\begin{table}
\begin{center}
\begin{tabular}{|c|c|c|}\hline
                      & $c_1$       & $c_2$ \\\hline
analytic              & $1.070$     & $0.11$    \\
static propagator     & $1.071(1)$  & $0.20(7)$ \\
Polyakov loop         & $1.070(3)$  & $0.17(14)$ \\
Reference\cite{lat99} & $1.070(3)$  & $0.14(4)$ \\\hline
\end{tabular}
\end{center}
\caption[]{Results for the static quark self-energy.}
\label{comp}
\end{table}

%%%%%%%%%%%%%%%%%%%%%%%%%%%%%%%%
\section{WILSON QUARK REST MASS}
%%%%%%%%%%%%%%%%%%%%%%%%%%%%%%%%
We now apply similar methods to the Wilson quark mass renormalization. In order to accommodate twisted boundary conditions consistently for propagating quarks, we enlarge the quark internal degrees of freedom to $N_{spin}\times N_{color}\times N_{smell}$ where the {\it smell} group is $SU(N_{smell}=3)$\cite{wohlert}. The boundary condition for the quark field reads,
\begin{equation}
\Psi(x+L_j\hat{e}_j)=\Omega_j\Psi(x)\Omega_j^{-1}e^{i\pi/N_{color}}
\end{equation}
where the phase factor ensures anti-periodicity. The momenta of the quarks are then given by,
\begin{equation}
p_i=\frac{\pi}{N_{color}L_i}(2n_i+1)
\end{equation}
\begin{equation}
n_x-n_y+n_z=mN_{color}
\end{equation}
where $m$ and $n_i$ are integers. The form of the constraint on $n_i$ is due to our particular choice of the twist matrix in the z-direction.

We have measured the ground state only, $(n_x=0,n_y=0,n_z=0)$, for which the {\it color-smell} wavefunction is proportional to the identity matrix\cite{wohlert}. The extrapolation to the infinite volume limit will then give the rest mass, $E(\vec{p}=\vec{0})$, 
\begin{equation}
M_1 = M_1^{[0]}+\alpha M_1^{[1]}+\alpha^2M_1^{[2]}+\alpha^3M_1^{[3]}+\cdots
\end{equation}
where $M_1^{[0]}=ln(1+am_0)$, $am_0=1/(2\kappa)-4$ and $M_1^{[1]}$ is calculated in Ref.~\cite{ask}. As explained in Ref.~\cite{hquark}, the parameters in the Wilson and clover actions should be taken as mass-dependent when the masses are large. Here we calculate the mass renormalization for $\kappa=0.1$. 

The ground state momenta was projected out both at the source and the sink. Only the spin components 0 and 2 (set equal to each other) were used in the simulation. 

\subsection{Fitting Procedure}
\begin{figure}
\begin{center}
\epsfxsize=2.5in
\epsfbox{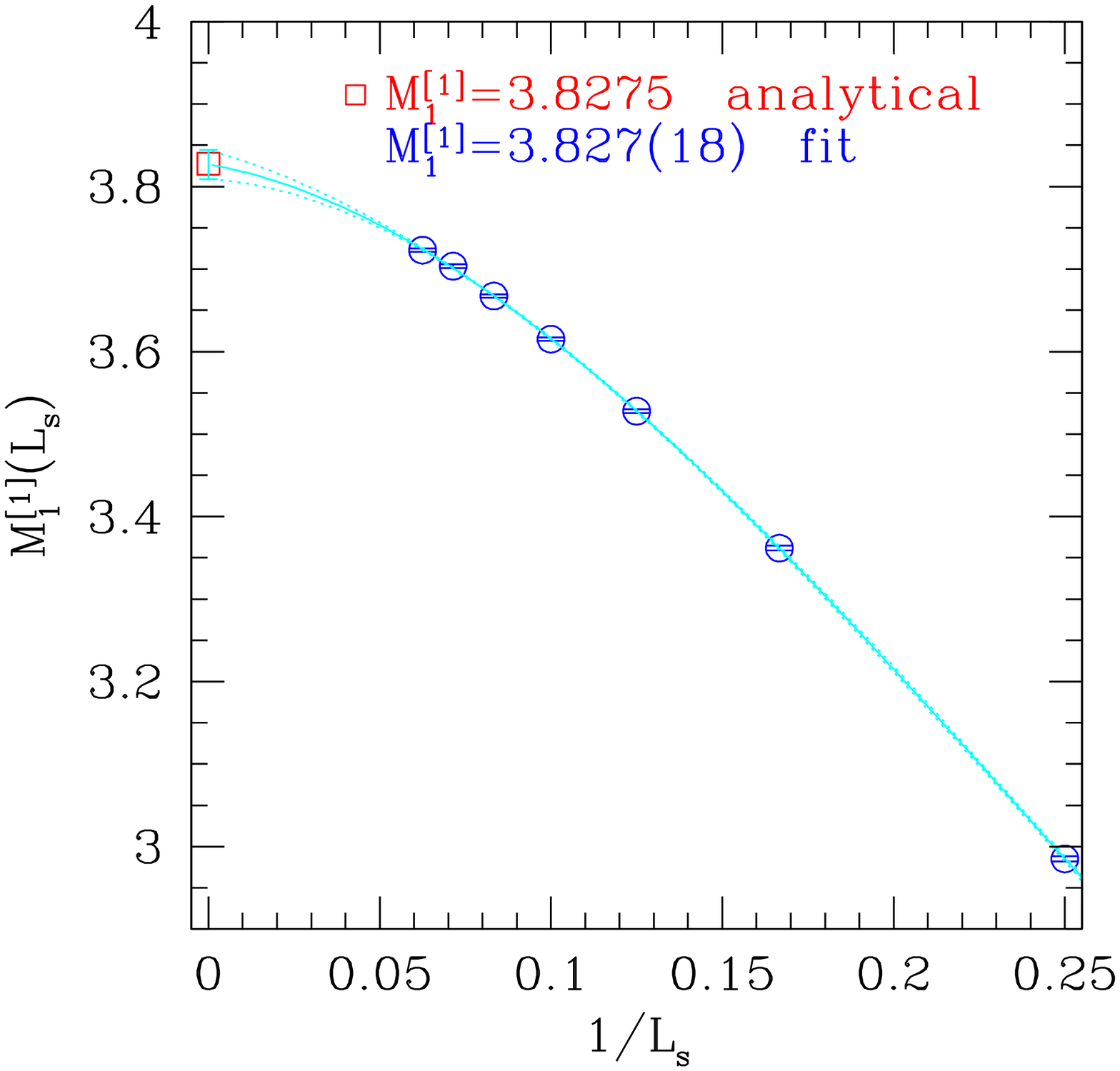}
\end{center}
\vspace{-12ex}
\caption[]{The one-loop coefficient as a function of the spatial lattice size.}
\label{W_c1}
\end{figure}

We have fitted the quark propagator to a $cosh$ starting at time-slice 3 for which we get a good $\chi^2$. The tree-level mass, $M_1^{[0]}(L)$, was then subtracted from the raw data so as to cancel some of the finite volume dependence of the rest mass. The extraction of the one-loop coefficient, $M_1^{[1]}$, and the two-loop coefficient, $M_1^{[2]}$, was done using the same procedure as the static quarks. However, we have found that a $1/L^2$ and $1/L^3$ term was required to get a good fit to the one-loop coefficient (see Fig.~\ref{W_c1}). The second order coefficient was extracted with a $1/L$ and $ln(L)/L$ term.

\begin{figure}
\begin{center}
\epsfxsize=2.5in \epsfbox{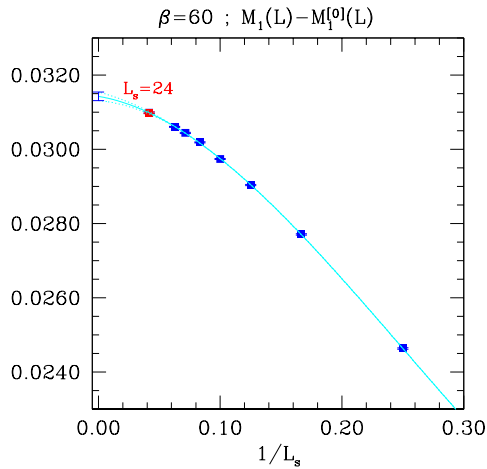}
\end{center}
\vspace{-12ex}
\caption[]{The infinite volume extrapolation of $M_1-M_1^{[0]}$ at $\beta=60$. Note that the point $L=24$ was not used in the fit.}
\label{b60}
\end{figure}

\begin{figure}
\begin{center}
\epsfxsize=2.5in \epsfbox{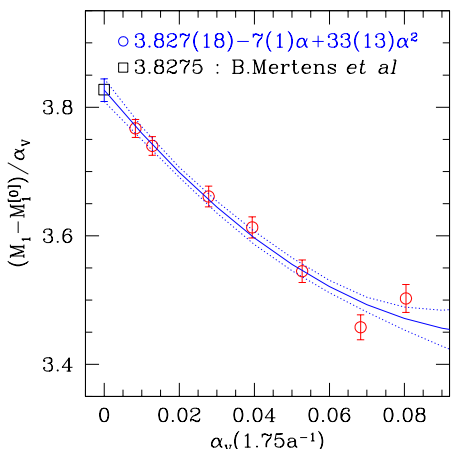}
\end{center}
\vspace{-12ex}
\caption[]{The fit to $(M_1-M_1^{[0]})/\alpha_V$ of the infinite volume extrapolated masses. ($q^\star=1.75a^{-1}$)}
\label{Malpha}
\end{figure}

As in the case of the static energy, we fit the masses in the reverse order to check for consistency. The infinite volume extrapolation is done with $1/L,1/L^2,1/L^3,ln(L)/L$ with the coefficient of $ln(L)/L$ fixed as before. We show one example in Fig.~\ref{b60}. The fit to a third order polynomial in $\alpha_V$ is shown in Fig.~\ref{Malpha} where the value of $q^\star$ was obtained in Ref.~\cite{ask}. The results are tabulated in Table~\ref{M1}.

\begin{figure}
\begin{center}
\epsfxsize=2.5in
\epsfbox{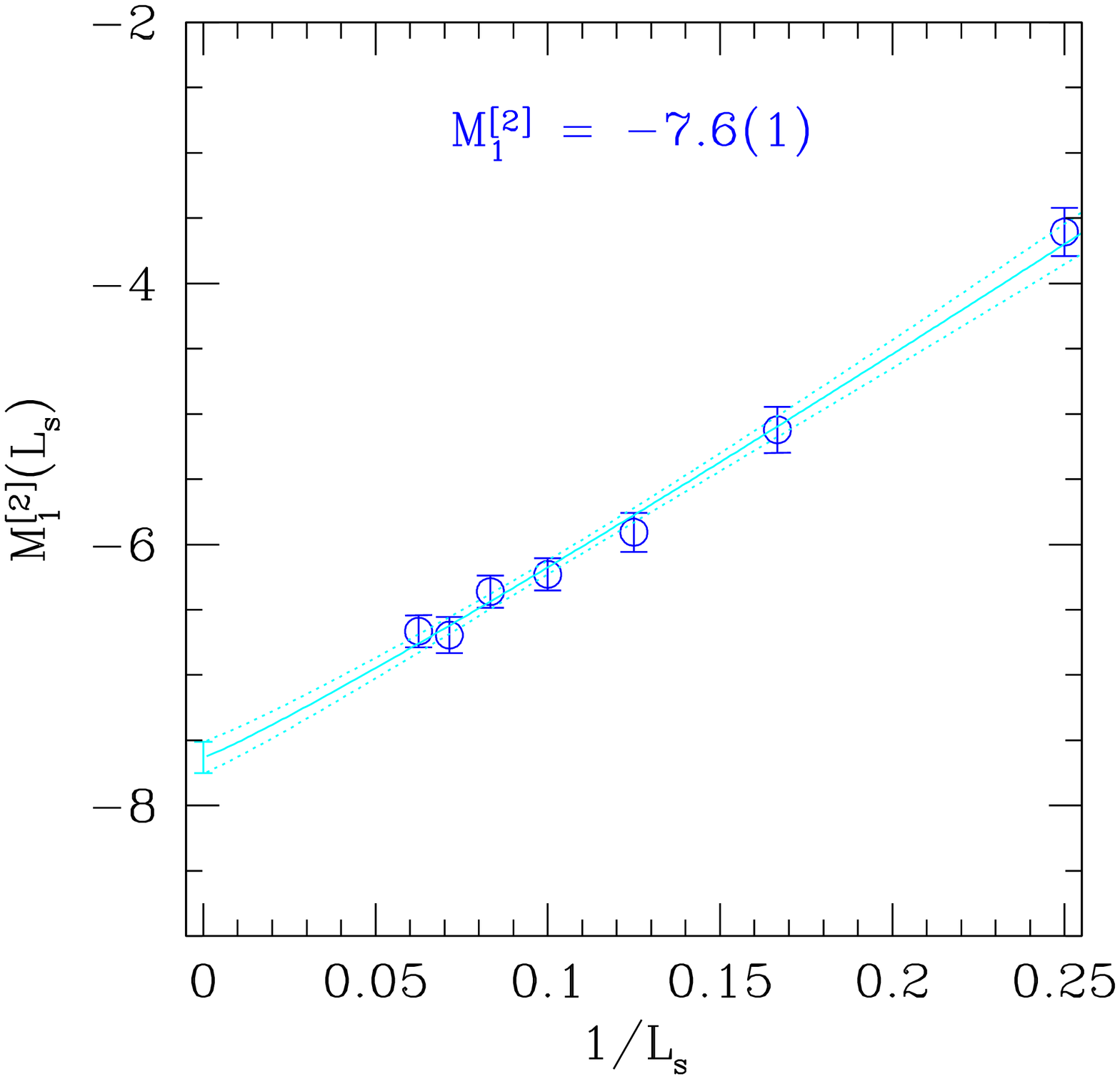}
\end{center}
\vspace{-12ex}
\caption[]{The two-loop coefficient as a function of the spatial lattice size.}
\label{W_c2}
\end{figure}

\begin{table}
\begin{center}
\begin{tabular}{|c|c|c|}\hline
                      & $M_1^{[1]}$ & $M_1^{[2]}$ \\\hline
analytic              & $3.8275$   & unknown    \\
$M^{[i]}(L)$ extrapolation& $3.827(18)$& $-7.6(1)$ \\
reverse order         & $3.827(18)$& $-7(1)$ \\\hline\hline
our estimate          & $3.83(2)$  & $-7(1)$ \\\hline
\end{tabular}
\end{center}
\caption[]{Results of the fit to the Wilson quark mass renormalization constants for $\kappa=0.10$}
\label{M1}
\end{table}

%%%%%%%%%%%%%%%%%
\section{SUMMARY}
%%%%%%%%%%%%%%%%%
It was shown that the one-loop Wilson quark mass renormalization constant is reproducable with a combined extrapolation and statistical error of $0.5\%$ from large $\beta$ Monte-Carlo simulations of the quark propagator. Twisted boundary conditions ensured a perturbative expansion around the trivial vacuum and the absence of zero modes. The gauge fixing procedure was checked to reproduce the static quark self-energy using both the Polyakov loop and the static quark propagator. 

Our preliminary result for the two-loop Wilson quark mass renormalization (for $\kappa=0.1$) is $-7(1)(0.5)$. We have a crude result for the three-loop term (see Fig.~\ref{Malpha}), but it is poorly determined and sensitive to the details of the fit procedure. For now, we simply take it as an estimate of the systematic error in the determination of the two-loop term. A more detailed study of the systematic errors is now in progress including the infinite volume extrapolation, dependence on $q^\star$ and truncation of the series. The heavy quark mass renormalization for improved actions, such as clover and NRQCD, are also currently under investigation.                                                                                                                        %%%%%%%%%%%%%%%%%%%%%%%%%%
\section*{ACKNOWLEDGEMENTS}
%%%%%%%%%%%%%%%%%%%%%%%%%%
I would like to thank A.~Kronfeld, P.~Lepage, P.~Mackenzie, N.~Shakespeare, J.~Simone and H.~Trottier for many useful discussions. Fermilab is operated by University Research Association, Inc. for the U.~S.~Department of Energy.

%%%%%%%%%%%%%%%%%%%%%%%%%%

\end{document}